\def\fej{
\magnification=1200
\hsize=6.5truein
\null
}
\def\br{\hfil\break} 
\def\df#1{{\tt #1 DEFINITION :}}
\def\pr#1{{\tt #1 PROPOSITION :}}
\def\noi{\noindent}

\def\diagram#1{{\normallineskip=8pt
    \normalbaselineskip=0pt \matrix{#1}}}

\def\rharr#1#2{\smash{\mathop{\hbox to .5in{\rightarrowfill}}
     \limits^{\scriptstyle#1}_{\scriptstyle#2}}}
\def\dvarr#1#2{\llap{$\scriptstyle #1$}\left\downarrow
  \vcenter to .5in{}\right.\rlap{$\scriptstyle #2$}}

\def\lharr#1#2{\smash{\mathop{\hbox to .5in{\leftarrowfill}}
     \limits^{\scriptstyle#1}_{\scriptstyle#2}}}
\def\uvarr#1#2{\llap{$\scriptstyle #1$}\left\uparrow
  \vcenter to .5in{}\right.\rlap{$\scriptstyle #2$}}
\def\ot{\otimes}

\fej
\vskip+1in
 
\bf 
\centerline{ON THE ROLE OF MASS IN THE}
\centerline{MATHEMATICAL STRUCTURE OF NEWTONIAN AND} 
\centerline{SPECIAL RELATIVISTIC MECHANICS}

\vskip+0.3in
\rm
\centerline{G\'abor Zsolt T\'oth\footnote{*}{tgzs@ludens.elte.hu}}
\centerline{Roland E\"otv\"os University, Faculty of Natural Sciences,
Budapest,} 
\centerline{Hungary} 
\vskip+0.1in

\bf
\centerline{Abstract}
\vskip+0.1in

{\parindent=.5in\narrower\sevenrm
We consider five-dimensional real linear spaces with a (otherwise well-known) linear action of the Galilei and the Poincare group on them, describe the geometry of these two spaces, and show, that these geometries comprise the notions of space-time, mass, momentum, force and physical dimensions in a natural way. In this way we geometrize the quantity of mass and integrate it together with space-time into two  geometries in a natural way, so that these geometries are perfectly suitable for underlying for the Newtonian and special relativistic mechanics of pointlike bodies. \par}

\vskip+0.3in

\centerline{I. INTRODUCTION}
\vskip+0.1in
\rm
\par
A classical mechanical theory begins usually with kinematics, i.e. with the
definition and description of some geometrical notions underlying the theory,
namely of space, time and motion.\br
The introduction of space-time by Einstein was a great step towards the
proper understanding of the geometrical structure of space and time, and
soon after that the well known nonrelativistic or Galilei space-time for
Newtonian mechanics was also  introduced by H. Weyl (see [1]). One can find
the definition of this space-time in many modern books, e.g. [2] [3] [4].\br
It is clear, that it fits very well to Newtonian mechanics. It is very
mysterious, however, that in order to be able to treat Newtonian dynamics
(not some generalized version of it) we have to introduce essentially one
(and only one) more quantity, namely the mass. It seems to us, that while
the structure of space-time is very well understood, this is not the case
with the step from kinematics to dynamics, although there are some
well-known   
results about the mass of free particles within the framework of
Hamiltonian mechanics and quantum theory. One usually says (in Newtonian
dynamics) that the mass is some parameter. One also refers to fundamental
experiments sometimes (e.g. [2]).\br
The main purpose of this paper is to offer an answer to this problem, i.e.
the problem of the role of mass in the Newtonian and in the special
relativistic mechanics of pointlike bodies. 
\br

Our answer is given as a reformulation of the elements of the two mechanical
theories. We did neither aspire
to treat the whole material of Newtonian mechanics and special
relativity, nor did we want to provide an
introduction. We restrict ourselves to those parts which we find necessary.
We assume that the reader is familiar with Newtonian
mechanics and special relativity and the abounding standard
mathematical notions. 
\hfil\break
\par

In section II. we treat Newtonian mechanics, in section III. special
relativistic mechanics in an analogous manner, and in section IV. we deal
with the relationship between the quantity which we call mass and the
quantity which is called mass in the literature (in particular in the
theorems about free particles and their relationship with the Galilei-group).
In the appendix we introduce some operations and notions which are used
throughout the text. \br

The fundamental object will be a geometrical structure denoted by $(V,G)$
in the formulation of both mechanical theories. This is similar to a
$G$-module, and consists of two parts:\br
-- a {\sl linear} space $V$\hfil\break
-- a subgroup $G$ of $GL(V)$, which we call symmetry group.\hfil\break

In particular, we shall have {\sl five-dimensional linear} spaces, and the
symmetry group will be the Galilei group and the Poincare group (whose action
will be {\sl linear}). We will see, that these spaces comprise the usual
quantities of mass, space, time, momentum, force, etc. so we shall call them
Newtonian and Einsteinian mechanical spaces. In this way mass will be
geometrized, and the classical abyss between kinematics and dynamics will be
lessened. \br

We will see, that it is possible to regard larger groups than the Galilei and
the Poincare group as symmetry groups of mechanics, we will offer an answer
to the question about the mathematical role or background of
the fact that there are three
independent dimensions (mass, length and period of time) in
Newtonian mechanics (and two in spec.rel.), and we will also see that the two
mechanical theories can be formulated without orientations (of space, time,
mass).\br

In our presentation the group theoretical point of view is favoured. This
will mean
that we shall  introduce various structures (so far as possible) as ones
determined by the symmetry group (so far as possible) and we shall distinguish
and appraise the various structures according to their invariance properties
(i.e. relationship to the symmetry group). This point of view is common in
modern physics (probably since Wigner), in geometry since the Erlangen
programme and in a major part of present day mathematics as well (through the
notion of category).\hfil
\vfill
\eject

{\bf
\centerline{II. NEWTONIAN MECHANICS}}
\vskip+0.2in

\noindent
{\bf 1. Definition of the Newtonian space}\hfil\break

\noindent
{\tt 2.1 DEFINITION: }\hfil\break
We call the following real Lie group the Galilei
group:\hfil
$$G^R=\left\{\pmatrix{O & v & x \cr
             0 & 1 & t \cr
             0 & 0 & 1 \cr} \in GL(R^3\times R\times R)\ |\ O\in SO(3),\ x,
             v\in R^3,
\ t\in R \right\}$$

\noindent
Let $V$ be a 5-dimensional real linear space, $G$ a subgroup of
$GL(V)$ so that there is a $b : V \to R^5 $ linear isomorphism for
which the map $i : GL(V) \to GL(R^5) ,\ \  g \mapsto bgb^{-1}$ establishes a
group isomorphism between $G$ and $G^R$.
\hfil\break
We call the pair $(V,G)$ an (unoriented) Newtonian mechanical
space, and we call a map $b$ satisfying the former condition an
inertial reference frame. Given two inertial reference frames $b_1, b_2$ we
call ${b_1}{b_2}^{-1}$ the map between the two reference frames.
If we have a certain $(V,G)$ specified,
then we call also $G$ Galilei group (or the Galilei group belonging to $V$).
The elements of $G$ act canonically on $V$ by linear automorphisms
$G \times V \to V,\ \ (g,v) \mapsto g(v)$ so the elements of $G$ are also
called Galilei transformations.
Obviously
$(R^5, G^R)$ is an example
of Newtonian mechanical spaces which we call coordinate space.
\hfil\break
We call two Newtonian mechanical spaces $(V_1, G_1) , (V_2, G_2)$
isomorphic if there exists a $b : V_1 \to V_2 $ linear isomorphism so
that the map  $i : GL(V_1) \to GL(V_2) ,\ \  g \mapsto bgb^{-1}$ establishes
a
group isomorphism between $G_1$  and $G_2$. The Newtonian
mechanical spaces constitute the objects of a category whose
morphisms are the isomorphisms defined just now. All objects of
this category are isomorphic. $\diamond$
\hfil\break

\noindent
$G^R$ is a very well-known form of the Galilei
group. The action of an element of $G^R$ looks like this:
\hfil\break
$$\pmatrix{O & v & a \cr
           0 & 1 & t \cr
           0 & 0 & 1 \cr} 
\pmatrix{x \cr
         y \cr
         z \cr}=
\pmatrix{Ox+vy+az \cr
         y+tz \cr
         z \cr}.$$
\break

\par 
\noindent
{\bf 2. Properties of the Newtonian space}\hfil\break
\noi

We describe some properties of the Newtonian mechanical
spaces now. For this purpose we assume that we are given a certain
Newtonian mechanical space $(V,G)$, which we can also think of as the proper
physical space.
\hfil\break

\par
\noindent
{\tt 2.2. PROPOSITION:}\br
The group of automorphisms of $(R^5, G^R)$ 
is a 13-dimensional real Lie group\hfil
 $${\bar G}^R=\left\{\pmatrix{A & a & b \cr
             0 & d & c \cr
             0 & 0 & e \cr} \in GL(R^3\times R\times R)\ |\ AA^{T}= n\cdot
             Id,\ n\in R^{+},\ d,e\in R\setminus \{ 0\},\right.$$
$$\left. \ a,\ b\in R^3\ \right\}$$

According to the definition the elements of ${\bar G}^R$ are the
transformations between inertial reference frames.
\hfil\break
Each element $\bar g$ of ${\bar G}^R$ can uniquely be written in the
form \hfil
$${\bar g} = cg, $$
where $c \in C^R$, $g \in G^R$, $C^R$ is a subgroup of ${\bar G}^R$
:\hfil
$$C^R=\left\{ \pmatrix{a\cdot Id & 0 & 0 \cr
            0 & b & 0 \cr
            0 & 0 & c \cr }\ \in GL(R^3\times R\times R\ ) |\ a,b,c\in
            R\setminus \{ 0\}    \right\}.$$
$G^R$ is an invariant subgroup of ${\bar G}^R$. $C^R$ is not
invariant, so ${\bar G}^R$ is the semidirect product of $C^R$ and $G^R$.
$\diamond$

\hfil\break
\noindent
{\tt 2.3. DEFINITION:} \hfil\break
Denoting the automorphism group of $(V,G)$ by $\bar G$, let $C = {\bar G} /
G$. $\diamond$ \hfil\break

\noi
$C$ is canonically isomorphic to $C^R$. $C$ is not invariant subgroup of
$\bar G$, but there is a class of conjugate subgroups of $\bar G$ isomorphic
to $C$. \hfil\break

\noindent
{\tt 2.4. PROPOSITION:}\br
The following subsets of $R^5 \equiv R^3\times R\times R$ are invariant under
the action of $G^R$ :\hfil\break
--- for all $m \in R$ the set $M_m^R = \{ (x,y,z) \in R^5\ |\ z=m \}$,
\hfil\break
--- for all $mt \in R$ the set $E_{mt}^R = \{ (x,y,z) \in R^5\ |\ y=mt,\ z=0
\}$,\hfil\break
--- for all $md \in R_0^{+}$ the set $S_{md}^R = \{ (x,y,z) \in R^5\ |\
\sqrt{< x,x >} = md ,\ y=z=0 \}$.\hfil\break
\noindent
($<,>$ means the usual scalar product on $R^3$.)\hfil\break

\noindent
The sets $M_m^R$, $m \in R$ are parallel 4-dimensional hyperplanes,\hfil
\break
the sets $E_{mt}^R$, $mt \in R$ are parallel 3-dimensional hyperplanes in
$M_0^R$,\hfil\break
the sets $S_{md}^R$, $md \in R_0^{+}$ are 2-dimensional similar spheres
around the $0$ contained in $E_0^R$.\hfil\break
(By 'similar' we mean that $S_x^R = (x/y) \cdot S_y^R$ for $x,y \in R, y\ne
0$.)\hfil\break

\noindent
The orbits of the action of $G^R$ are : \hfil\break
$M_m^R\ ,\ \ m \in R\setminus \{ 0\}$ \hfil\break
$E_{mt}^R\ , \ \ mt \in R\setminus \{ 0\}$ \hfil\break
$S_{md}^R\ , \ \ md\in R_0^{+}$.\hfil\break

\noindent
We see from this that $V$ also decomposes to invariant hyperplanes 
and spheres uniquely in the way described above. By sphere we mean the level
set of some positive definite quadratic form. This decomposition can be
obtained by pulling back the one of $R^5$ to $V$ with any arbitrarily chosen
inertial reference frame. $\diamond$
 \hfil\break

\noindent
Let us introduce the following notation for the components of $V$:\hfil\break
$M_{m(v)}$ denotes the 4-dimensional hyperplane which contains $v \in V$,
\hfil\break
$E_{mt(v)}$ denotes the 3-dimensional hyperplane in $M_0$ which contains
$v$, where $M_0 = M_{m(0)}$,\hfil\break
$S_{md(v)}$ denotes the sphere in $E_0$ which contains $v$, where $E_0 =
E_{mt(0)}$.\hfil\break

\noindent 
{\tt 2.5. PROPOSITION:} The subgroup of $GL(V)$ which preserves the above
decomposition of $V$ is $\bar G$. The connected subgroup under whose action
the components of the above decomposition are invariant is the Galilei group
$G$.
 $\diamond$
\hfil\break

This means, that the Newtonian mechanical space could have been defined in
terms of the above  decomposition. \hfil\break

Let us denote \hfil\break
the set of all 4-dimensional hyperplanes of the decomposition of $V$ by
$[kg]$, \hfil\break
the set of all 3-dimensional hyperplanes of the decomposition of $V$ by
$[kgs]$,
\hfil\break
the set of all spheres of the decomposition of $V$ by $[kgm]^{+}$.\hfil
\break

The reason for this notation will be clear later. 
$[kg]$ has an 1-dimensional  linear space structure : $[kg] \equiv V/M_0$.
$[kgs] \equiv M_0/E_0$ is a 1-dimensional linear space as well and
$[kgm]^{+}$ is the positive part of a 1-dimensional oriented linear space
(multiplication with a real number can be defined through a representative
and addition can be defined using two parallel representatives). Note, that
$[kg]$ and $[kgs]$ are not oriented.\hfil\break

The group $\bar G$ acts on $[kg]\times [kgs]\times [kgm]$ by linear
isomorphisms. The kernel of this action is the subgroup $GP$ (a group with
two topological components) of $\bar G$ under the action of which the
components of $V$ are invariant. (This group acts on $E_0$ effectively as an
$O(3)$.) It is the (multiplicative) group $\bar G / GP\equiv R^{+}\times
(R\setminus \{0\})\times (R\setminus \{0\})$  which acts effectively on
$[kg]\times [kgs]\times [kgm]$.
(Note, that $C\equiv (R\setminus \{0\})\times (R\setminus \{0\})\times
(R\setminus \{0\})$ and ${\bar G}/GP \equiv C/(\{ -1, 1\} \times \{ 1\}
\times\{ 1\})$.) \hfil\break

\noindent
{\tt 2.6. DEFINITION:}\hfil\break
We introduce the following ($\bar G$- equivariant) maps:\hfil\break

\noindent
$m : V \to [kg],\  v\mapsto M_{m(v)}$\hfil\break
$mt : M_0 \to [kgs],\  v \mapsto E_{mt(v)}$\hfil\break
$md : E_0 \to [kgm]_0^{+},\  v\mapsto S_{md(v)}$.$\diamond$
\hfil\break 

\noindent
The first two ones are linear maps.\hfil\break

\noindent
In the remaining part we need some operations with one-dimensional linear
spaces. These are introduced in the appendix, so the reader is 
advised to read it before further advance.\br

\noindent
It is clear, that $md$ determines a positive definite bilinear map\hfil
 $$<,>\ :\ E_0 \times E_0 \to [kgm]^2,\ \  (v,w)\mapsto 1/4\cdot
 ((md(v+w))^2-(md(v-w))^2),$$  which we call the (generalized) Euclidean
 scalar product, which  also determines a tensor $g \in Sym^2 (E_0^{*})
 \otimes [kgm]^2 \equiv Hom(Sym^2(E_0), [kgm]^2)$.\hfil\break

The situation is the following now: we have two linear maps and a symmetric
bilinear one, which differ from ordinary linear and bilinear forms only in
that they are not real but one-dimensional linear space valued. We shall
call such forms generalized forms.\hfil\break
Another (equivalent) way to describe the situation is the following:  the
decomposition of $V$ determines 1-dimensional subspaces in $Hom(V, R)$ and
in $Hom(M_0, R)$ and a half of a linear space in $Hom(Sym^2(E_0), R)$  (this
half is the positive definite half), in other words we have linear forms
determined up to scalars, i.e we have a kind of conformal structure on $V$.
(The mentioned spaces are just $[kg]^{-1}, [kgs]^{-1}, [kgm]^{-2}$, they are
embedded into $Hom(V, R)$, $Hom(M_0, R)$ and $Hom(Sym(E_0), R)$ by the
transpose maps of $m$, $mt$ and $g$.)\hfil\break

Note, that every inertial reference frame induces  linear isomorphisms of
$[kg]$, $[kgm]$ and $[kgs]$ with $R$. These isomorphisms are the same for two
inertial reference frames which are related by a map in $GP^R$. \hfil\break
All spaces obtained from $V$ by multiplication or division by some power of
$[kg]$, $[kgm]$ or $[kgs]$ have a unique Newtonian structure on them
isomorphic to the one on $V$ up to real or positive real numbers, and an
inertial reference frame of $V$ determines unique inertial reference frames
on them.
\hfil\break

Let us now introduce the following linear spaces:\hfil\break

\noindent
{\tt 2.7. DEFINITION:} \hfil\break
$[m]:=|([kgm]/[kg])|,$\hfil\break
$[s]:=[kgs]/[kg].$\hfil\break
We call $[kg]$ the measure line of mass,\br
$[m]$ the measure line of distance and\br
$[s]$ the measure line of periods of time. $\diamond$
\hfil\break

The notation $[m]$ comes from 'meter', $[s]$ comes from 'second' and $[kgm]$
comes from 'kilogram'.\br

\noindent
We specify now some groups related to $G$:\hfil\break

\noindent
{\tt 2.8. DEFINITION:}\hfil\break 
--- ${\cal T}^4$ : group of parallel translations, the kernel of the
homomorphism $G \to G\lceil M_0$\footnote{*}{$\lceil$ denotes the restriction
of a map.}, a 4-dimensional Abelian Lie group, canonically isomorphic to the
additive group of $M_0/[kg]$,\hfil\break
--- $SO{\cal B}\equiv G\lceil M_0$ : homogenous Galilei group, 6-dimensional,
\hfil\break
--- $\cal B$ : group of velocity transformations (i.e. Galilei boosts), the
kernel of the homomorphism $G\lceil M_0 \to G\lceil E_0$. It is a
3-dimensional Abelian Lie group, canonically isomorphic to the additive group
of $E_0/[kgs]$,\hfil\break
--- $SO(g) \equiv G\lceil E_0$: group of rotations, a simple Lie group
(isomorphic to  $SO(3)$),\hfil\break
--- ${\cal BT}^4$ : the kernel of the homomorphism $G\to G\lceil E_0$,
nilpotent 7-dimensional Lie group,\hfil\break
--- ${\cal T}^3\equiv  E_0/[kg]$: group of spacelike translations,
canonically isomorphic to the additive group of $E_0/[kg]$. $\diamond$
 \hfil\break

Consider now the following figures and sets:\hfil\break

\noindent
{\tt 2.9. DEFINITION:}\hfil\break
$M=\{$ lines in $V$ containing 0 and not lying in $M_0 \}$,\hfil\break
$T=\{$ 4-dimensional subspaces in $V$ containing $E_0 \}\setminus M_0 $,
\hfil\break
$E^3(t)=\{ e\in M | e \subset t\}$ for all $t\in T$.\hfil\break
$\bar G$ acts on both $M$ and $T$. $\diamond$\hfil\break

\noindent
{\tt 2.10. PROPOSITION:} $M$ is a 4-dimensional affine  space over
$M_0/[kg]$, $T$ is a 1-dimensional affine space over $[s]$. $G$ acts on $T$
by the additive group of $[s]$ effectively. $\diamond$\hfil\break

\noindent
{\tt 2.11. DEFINITION:} \hfil\break
 We denote the additive group of $[s]$ by ${\cal T}^1$ and call it the
 group of time translations. $\diamond$ \hfil\break

\noindent
{\tt 2.12. DEFINITION:}\hfil\break
We denote the kernel of the homomorphism $G \to {\cal T}^1$ by $SO{\cal
BT}^3$. Let ${\cal BT}^3 = SO{\cal BT}^3 \cap {\cal BT}^4$. $\diamond$
 \hfil\break

\noindent
{\tt 2.13. PROPOSITION:} (The normal subgroups of $G$) We have all normal
Lie subgroups of $G$ now, these are: ${\cal T}^3,\ {\cal T}^4,\ {\cal BT}^3,\
{\cal BT}^4,\ SO{\cal BT}^3$. The corresponding factor groups are: $SO{\cal
BT}^1,\ SO{\cal B},\ SO{\cal T}^1,\ SO(g),\ {\cal T}^1$. In figure 1. and 2.
we can see the net of normal Lie subgroups of $G$ and the dual net of factor
groups. The arrows indicate canonical inclusions and canonical homomorphisms.
$\diamond$ \hfil\break

For all ${t\in T}$ the $E^3(t)$ is an affine subspace of $M$ over $E_0/[kg]$,
so there is a (conformal) Euclidean structure on every $E^3(t)$.\hfil\break

We introduce the following maps and notations:\hfil\break

\noindent
{\tt 2.14. DEFINITION:}\hfil\break 
$u : V\setminus M_0 \to M,\  p \mapsto \{$ the line in $M$ containing $p
\}$\hfil\break
$\tau : V\setminus M_0 \to T,\  p\mapsto \{$ the 4-dimensional subspace in
$T$ containing $p \}$\hfil\break
$d_t : E^3(t)\times E^3(t) \to [m],\ \ \  (x,y) \mapsto ||x-y|| = md(x-y).
$    \hfil\break

Note, that $u(p) = p/m(p)$. $M$ is canonically embedded into $V/[kg]$ and as
a subset it is determined by the property $(m/[kg])(M) = 1$.
$\diamond$
 \hfil\break

We introduce now the following names: \hfil\break

\noindent
$m$ : mass (evaluation function)\br
$M$ : Galilei space-time, the elements of which are called events,\hfil\break
$T$ : time line, the elements of which are called points of time,\hfil\break
$E^3(t)$ : synchronous space at $t\in T$,\hfil\break
$\tau$ : time (evaluation function),\hfil\break
$u$ : place (in the space-time, evaluation function), \hfil\break
$d_t$ : synchronous distance function at $t \in T$.\hfil\break

$M$ with $\tau$ and the $d_t$-s (or equivalently $M$ with the affine
$G$-action on it) forms a usual Galilei spacetime, which was introduced by H.
Weyl [1] ( see also [2], [3], [4]). Note, that $\tau$ is a synchronization of
$M$ and makes $M$ a bundle over $T$.\hfil\break
Note also that $V\setminus M_0\ =\ M\times ([kg]\setminus \{ 0\} )$, where
the projections are $m$ and $u$. \hfil\break

\noindent
{\tt 2.15. DEFINITION: } \hfil\break
We call $(V,G)$ an oriented Newtonian space, if its structure is supplemented
by an orientation of $[kg]$, $[kgs]$ and $E_0$. (This has a smaller
automorphism group.) $\diamond$\hfil\break

If a Newtonian space is oriented, we can speak of future and past, positive
and negative mass, and vectorial product in $E_0$, for example.\br

{\bf
\leftline{3. The Newtonian particle; momentum and force  }}
\vskip+0.1in

\noindent
{\tt 2.16. DEFINITION:}\hfil\break
A Newtonian pointlike particle is a function $f : I \to V $, which has the
following properties:\hfil\break
--- $I$ is a closed interval of $T$,\hfil\break
--- $m\circ f$ is a constant function, this constant is called the mass of
the particle (and denoted by $m(f)$),\hfil\break
--- $t\circ f = id_I$ (i.e. $f$ is a natural parametrization of its range).
$\diamond$\hfil\break

One can define the category of Newtonian particles, the objects of which are
the triplets $(V,G,f)$, where $(V,G)$ is a Newtonian space. A particle can be
regarded as an additional structure on a Newtonian space.\hfil\break

\noindent
{\tt 2.17. DEFINITION:}
We call $p= f'$ the four-momentum function of $f$. \hfil\break
Two properties of $p$ are: $Ran(p) \subset M_0/[s],\ \  mt \circ p \equiv
m(f)$ \hfil\break

We call $v = p/m(f) = (u\circ f)'$  the four-velocity function of $f$. It
satisfies the following: $ Ran(v) \subset M_0/[kgs],\ \  mt\circ v \equiv 1,
\ \  p = m(f)\cdot v$.\hfil\break

We call $F = f''$ the force acting on the particle $f$. It has the following
properties: $Ran F \subset E^3/[s]^2, F = m(f)\cdot v' = m(f)\cdot  {(u\circ
f)}'$.\hfil\break

We call $a = F/m(f) = v' = {(u\circ f)}''$ the acceleration function of the
particle. For the acceleration and force we have now\hfil
$$F = m(f)a= p'$$
$\diamond$
\hfil\break

Note, that in general it is $SO{\cal BT}^1$ which acts on the four-velocity
functions and four-momentum functions, and it is $SO{\cal T}^1$ generally
which acts on forces and acceleration functions. $SO{\cal B}$ acts on the
range of four-velocity functions and $SO(g)$ acts on the range of forces.
\hfil\break

\noi
{\bf 4. Force field, equation of motion for one body}\hfil\break

\noindent
{\tt 2.18. DEFINITION:}\hfil\break
A function $F : V \times M_0/[s] \to E^3/[s]^2$ with an open
domain\footnote{*}{This is not the weakest satisfactory condition for the
domain.} is called a force field. The differential equation\hfil
$$f'' = F \circ ( f, f' )$$
for a particle $f$ is called Newton's equation of motion, and the solutions
of this equation are called the particles determined by the force field in
question.  An initial value must satisfy the following: $m(f(t_0)) = m_0\ne
0,\  m_0 = mt( f'(t_0))$ . $\diamond$
 \hfil\break

One can define the category of Newtonian force fields or Newtonian mechanical
systems (the objects of which are the triplets $(V,G,F))$. The automorphism
group of such an object can be called the dynamical symmetry of the system.
The only force field which is invariant under the Galilei group is the zero
field, which determines free particles.\hfil\break

\noindent
{\tt 2.19. DEFINITION:}\hfil\break
Let $f_i, i=1..N$ be N particles. We introduce their \hfil\break
center of mass : $f = \sum_{n=1}^{N} f_i$,  \hfil\break
total momentum : $f'$,\hfil\break
internal angular momentum : $J = \sum_{i<j} {{m_jf_i-m_if_j}\over m_i+m_j}
\wedge ( {f_i\over m_i} -{f_j\over m_j})'.$ $\diamond$
\hfil\break

\noi
{\bf 5. The space of four-velocities}\br

\noindent
\df {2.20.} \br
Let $V(1) = \{$ the lines in $M_0$ containing 0, not lying in $E_0\ \} \equiv
\{\ v \in M_0/[kgs],$\br
$mt(v)=1\ \} .$\br
$V(1)$ is the space from which the four-velocity functions take their values.
$\diamond$\br

\noindent
\pr {2.21.} \br
$V(1)$ is a 3-dimensional Euclidean affine space over $E_0/[kgs]$.\br
$M_0\setminus E_0 = V(1)\times ([kgs]\setminus \{ 0\} ), $ where the
projections are $id_{M_0}/mt$ and $mt$.  \br
$\diamond$\br

\noindent
{\bf 6. Effects of fixing a reference frame}\hfil\br

Now we mention a few important additional structures on $(V,G)$ which arise
when a particular inertial reference frame is chosen.\br

A certain timelike line in $M$ is determined, which is called the origin (of
the reference frame), $E_0$ gets an oriented orthogonal basis and the three
axes of space, $M$ gets a direct product structure : $M=(E_0/[kg])\times
[s]$, $V_1$ gets a zero and is mapped to $E_0/[kgs]$, so the kinetic energy
of a particle can be defined, $[kg]$, $[kgs]$, $[kgm]$ get units (and
orientation), and the factor groups listed in 2.13.  and $C$ obtain
well-defined monomorphisms into $\bar G$, $\bar G$ obtains a parametrization
and its Lie-algebra a basis. (Many other things could be mentioned, for a more
detailed description see e.g. [4].) \br

If a mass $m$, a position three-vector $x$ and a point of time $t$ is given
with respect to some reference frame, then the corresponding vector in
$R^5\equiv R^3\times R\times R$ is\hfil
$$(mx,mt,m).$$
The coordinate form of the points of $M$ is \hfil
$$(x,t,1),$$
of the points of $V(1)$ is\hfil
$$(v,1,0),$$
and the functions $m$, $mt$, $md$, $\tau$, $d_t$, $u$ have the following
forms:\hfil
$$m : (mx,mt,m) \mapsto m,$$
$$mt : (mx,mt,0) \mapsto mt,$$
$$md : (mx,0,0) \mapsto ||mx||,$$
$$\tau : (mx,mt,m) \mapsto t,$$
$$d_t : ((m_1x,m_1t,m_1),(m_2y,m_2t,m_2)) \mapsto ||x-y||,$$
$$u : (mx,mt,m) \mapsto (x,t,1).$$
If $f(t)=(mx(t),mt,m)$ is a particle, then\hfil
$$p(t)=(mx'(t),m,0),$$
$$F(t)=(mx''(t),0,0),$$
$$v(t)=(x'(t),1,0),$$
$$a(t)=(x''(t),0,0).$$

\vfill
\eject

{\bf \hfill III. SPECIAL RELATIVITY}\hfill\break

\noindent
{\bf 1. Einsteinian space}\hfil\br

\noindent
{\tt 3.1. DEFINITION:} \hfil\break
We call the following 10-dimensional Lie-group the Poincare group:\hfil

$$P^R = \left\{\pmatrix{L & x \cr
             0 & 1 \cr}
              \in GL(R^4 \times R) | L\in SO^{+}(3,1), x\in R^4 \right\}$$
$\diamond$
\hfil\break

We define the category of Einsteinian mechanical spaces by replacing the
Galilei group with the Poincare group in Definition 1.1. \hfil\break

\noindent
{\bf 2. Properties of the Einsteinian space}\br

For the description of the properties of Einsteinian mechanical spaces we
assume, that we are given a certain one: $(V,P)$.\br

\noindent
{\tt 3.2. PROPOSITION:}\br
The group of automorphisms of $(R^5, P^R)$ is a 12-dimensional Lie
group:\hfil
$${\bar P}^R = \left\{\pmatrix{A & a \cr
             0 & b \cr}
              \in GL(R^4 \times R) | A=n\cdot L,\ L\in SO^{+}(3,1), a\in R^4,
              \right.$$
$$\left. n,b\in R\setminus \{ 0\} \right\}.$$

\hfil\break
The elements of ${\bar P}^R$ are again the transformations between inertial
reference frames. Each element $\bar p$ of ${\bar P}^R$ can uniquely be
written in the form\hfil
$${\bar p} = cp,$$
where $p \in P^R$, $c\in C^R$ and $C^R$ is a subgroup of ${\bar P}^R$: \hfil
$$C^R=\left\{ \pmatrix{a\cdot Id & 0 \cr
                       0 & b \cr }\ \in GL(R^4 \times R\ ) |\ a,b\in R\setminus
                       \{ 0\}    \right\},$$

\noi
and $P^R \cap C^R = \{ Id_{R^5} \}$. $P^R$ is an invariant subgroup of
${\bar P}^R$, $C^R$ is not. $\diamond$
\hfil\break

Denoting the automorphism group of $(V,P)$ by $\bar P$ we define $C = {\bar
P}/P$, this is isomorphic to $C^R$. \hfil\break

\noindent
{\tt 3.3. PROPOSITION: } The following subsets of $R^5\equiv R^4\times R$ are
invariant under the action of $P^R$:\hfil\break
\noindent
--- for every $m\in R$  the set $M_m^R = \{ (x,z) \in R^5\ |\ z=m \}
$,\hfil\break
--- for every $mt\in R^{+}$ the set $H_{mt}^R = \{ (x,z)\in R^5\ |\ z=0,\ <x,
x> = (mt)^2\cdot (-1),\ \ x_4>0\ \}$,\hfil\break
--- for every $mt\in R^{-}$ the set $H_{mt}^R = \{ (x,z)\in R^5\ |\ z=0,\ <x,
x> = (mt)^2\cdot (-1),\ \ x_4<0\ \}$,\hfil\break
--- for every $md\in R^{+}$ the set $S_{md}^R = \{ (x,z)\in R^5\ |\ z=0,\ <x,
x> = (md)^2\ \} $,\hfil\break
--- the set $L^R = \{ (x,z)\in R^5\ |\ z=0,\ <x, x> = 0\ \}$\hfil\break
--- the set $S_0^R = \{\ 0\ \}.$ \hfil\break

\noi
Here $<,>$ is the standard product on $R^4$ with signature $(+++-)$.\br

\noindent
The sets $M_m^R$ are parallel 4-dimensional hyperplanes of $R^5$, \hfil\break
the sets $H_{mt}^R$ are similar connected components of 3-dimensional
two-sheeted hyperquadrics in $M_0^R$,\hfil\break
the sets $S_{md}^R$ are similar 3-dimensional one-sheeted hyperquadrics in
$M_0^R$,\hfil\break
$L^R$ is the 3-dimensional light cone in
$M_0^R$,\hfil\break
$S_0^R$ is just the 0 in $R^5$.\hfil\break

The orbits of the action of $P^R$ are:\br
$M_m^R,\  m\in R\setminus \{ 0\}$,\br
$H_{mt}^R,\ mt \in R\setminus \{ 0\}$,\br
$S_{md}^R,\ md \in R_0^{+}$, and\br
$L^R$. \hfil\break

Thus $V$ also decomposes uniquely to  invariant hyperplanes and hyperquadrics
and a cone. This decomposition can be obtained by pulling back the one of
$R^5$ to $V$ by any inertial reference frame. $\diamond$
\hfil\break

We introduce the following notation for the components of $V$:\hfil\break
$M_{m(v)}$ denotes the 4-dimensional hyperplane which contains $v$,
\hfil\break
$H_{mt(v)}$ denotes the connected component of the 3-dimensional two-sheeted
hyperquadric which contains $v$, where $v \in M_0\equiv M_{m(0)}$, \hfil\break
$S_{md(v)}$ denotes the 3-dimensional one-sheeted hyperquadric which contains
$v$, where $v \in M_0$,\hfil\break
$L$ denotes the light cone,\hfil\break
$S_0 = \{ 0 \}$. \hfil\break 

The analogue of Proposition 1.3. holds, so the Einsteinian structure could
have been defined in terms of the above decomposition.\hfil\break
We see, that we have a kind of conformal structure again.\hfil\break

Let us denote the set of all 4-dimensional hyperplanes of the decomposition
of $V$ by $[kg]$, \hfil\break
the set of the $H$ -s and $S_0$  by $[kgs]$, \hfil\break
the set of the $S$ -s and $S_0$  by $[kgm]_0^{+}$. \hfil\break
\noi
$[kg] = V/M_0\ $, $[kgs]$ is an unoriented 1-dimensional linear space
(multiplication and addition can be defined by using suitable representatives)
and $[kgm]_0^{+}$ is the nonnegative half of an oriented 1-dimensional linear
space $[kgm]$. We introduce further notations and maps:\hfil\break

\noindent
{\tt 3.4. DEFINITION:}\hfil\break
$[m]=|[kgm]/[kg]|$ : the measure line of distances,\hfil\break
$[s]=[kgs]/[kg]$ : the measure line of time periods,\hfil\break
$[kg]$ : the measure line of mass. \hfil\break
\noi
$m : V \to [kg],\ v \mapsto M_{m(v)}$\hfil\break
$mt : M_0 \to [kgs],\ v \mapsto H_{mt(v)}$\hfil\break
$md : M_0 \to [kgm]_0^{+},\ v \mapsto S_{md(v)}.$ $\diamond$
\hfil\break

\noindent
{\bf 3. The velocity of light}\br

The light cone in $M_0$ determines Lorentzian quadratic forms on $M_0$ up to
nonzero real factor, so by choosing one of them, say $l$, we can define two
maps, $c_1,\ c_2$ from $[kgm]$ to $[kgs]$ (i.e. two elements of $[kgm]/[kgs])$
as follows:\hfil

$$ c_1 ( x ) = l^{-1} ( (-1)\cdot l(x)),\ \ \ \ \ x \in [kgs]_1,$$
$$ c_2 ( x ) = l^{-1} ( (-1)\cdot l(x)),\ \ \ \ \ x \in [kgs]_2,$$
\noi
where $[kgs]_1$ and $[kgs]_2$ denote the two halves of $[kgs]$.
$c_1$ and $c_2$ should be extended linearly to the whole $[kgs]$. Then $c_1=
-c_2\ $.
$c_1$ and $c_2$ are independent of the choice of $l$. \hfil\break

An other way to define these elements of $[m]/[s]$ is to choose a
2-dimensional subspace in $M_0$ which has nonzero intersection with the
elements of $[kgs]$, i.e. Lorentzian. The light cone intersects this plane in
two lines (which intersect each other in 0). These lines determine reflections
which map the intersections of the elements of $[kgs]$ and $[kgm]^{+}$ with
the plane  into each other. Two maps can be obtained in this way (after
linear extensions), and they are independent of the choice of the
2-dimensional plane and are identical to $c_1$ and $c_2$. \hfil\break

\noindent
{\tt 3.5. DEFINITION:}\hfil\break
$c = |c_1| = |c_2| \in |[m/s]|$ is called the velocity of light. (This name
originates from electrodynamics.) $\diamond$
\hfil\break

(The velocity of light is -- obviously -- invariant under the action of $\bar
P$.) According to the definitions the velocity of light equals 1 in every
inertial reference frames. The vector space $[m]/[s]$ can be identified with
$R$ algebraically. The elements of $[m]$ and $[s]$ can be distinguished by
their geometrical meaning, however. \hfil\break

We have now the following maps on $V$:\br
\noi
$m : V \to [kg],$\br
$<,> : M_0\times M_0 \to [kgm]\ \ $ or $\ \  |[kgs]|.$\br

($m$ is already defined, $<,>$ is the generalized Lorentzian form.)\hfil\break
\noi
We shall give a description of the subgroups of $P$ now.\hfil\break

\noindent
\pr {3.6.} \br
The net of normal Lie subgroups and its dual net is the following: see fig.
3., 4.\hfil\break

\noi
${\cal T}^4$ is the kernel of the homomorphism $P \to P\lceil {M_0}$ ,
equivalent to the additive group of $M_0/[s],$\hfil\break
${\cal L} = P\lceil {M_0}$, it is called the (homogenous) Lorentz group.
\hfil\break

\noi
Other important subgroups of $\cal L$ :\hfil\break
\noi
--- stabilizers of timelike vectors in $M_0$ ; these are conjugate subgroups,
each of them is the special orthogonal group of the orthogonal space of the
stabilized vector,\hfil\break
--- stabilizers of spacelike vectors in $M_0$ ; these are conjugate subgroups
isomorphic to $SO(2,1)$, i.e. each of them is the special orthogonal group of
the orthogonal space of the stabilized vector,\hfil\break
--- stabilizers of lightlike vectors in $M_0$ ; these are conjugate subgroups,
isomorphic to the group (called Euclidean group) :\hfil
$$\left\{\pmatrix{O & 0 \cr
             a & 1 \cr}
              \in GL(R^2 \times R) | O\in SO(2), a\in R^2,\right\}.$$
(This is the same representation as the one we really have). Each of these
groups act on the orthogonal space of the stabilized vector. \hfil\break
--- groups of boosts: these are the groups  which preserve the orthogonal
decompositions $B_1\oplus B_2$ of $M_0$, where the components are nonsingular
subspaces and the action of which on the spacelike part of the decomposition
is the identity. The boosts form conjugate subgroups which are isomorphic to
the additive group of $R$.\hfil\break

\noindent 
{\tt 3.7. DEFINITION:} \hfil\break
$M=\{$ the lines in $V$ containing 0 and not lying in $M_0\ \}$ $\diamond$
\hfil\break

$M$ is a 4-dimensional affine space over ${\cal T}^4$, ${\cal T}^4\equiv
M_0/[kg]$. The stability groups of the points of $M$ are isomorphic to $\cal
L$.\hfil\break

\noindent
\df {3.8.} \br
We introduce the following map:\hfil\break
$u: V\setminus M_0 \to M,\ \  p \mapsto \{ $ the line in $M$ containing $p
\}$.
 $\diamond$\hfil\break

$M$, being an affine space over ${\cal T}^4$, carries a further structure:
the distance function \hfil
$$d: M\times M \to [m],\ \  (a,b) \mapsto ||(a-b)||$$.

\noindent
\df {3.9} \br
We introduce the following names\br

\noi
$m$ : mass (evaluation function)\br
$M$ : spacetime\br
$u$ : place (in spacetime, evaluation function)\br
$d$ : distance (in spacetime).
$\diamond$\br

$P$ acts on $M$ (obviously). $M$ with the affine structure and distance map
(or equivalently with the action of $P$) is a relativistic space-time or a
Minkowskian space (not vector space) in the usual sense. The elements of $M$
are called events. $\diamond$
\hfil\break

The spaces $M_m,\ \  m\in [kg], \ m\ne 0$ are Minkowskian spaces as well, and
their isomorphisms with each other arise from the action of the elements of
${\bar P}$ on $V$. (Note that the distance function is not real valued here,
but the elements of $\bar P$  do act on its values nontrivially!) In other
words, $d_{m_2}\circ p\times p = {\tilde p}\circ d_{m_1}$, where $p\in \bar
P$, $p(M_{m_1})=M_{m_2}$ and $d_m=m\cdot d.$\hfil\break
 
\noindent
{\bf 4. Particles, momentum, velocity, force}\br

\noindent
{\tt 3.10. DEFINITION: }\hfil\break
An Einsteinian pointlike body is described by a function $f : I \to V$ which
satisfies the following:\hfil\break
--- $I$ is a closed interval of $[s]$,\br
--- it is contained by some $M_m$ for some $m\in [kg], m \ne 0$, which is
called the mass of $f$, denoted by ($m(f)$),\hfil\break
--- $||\dot f|| \equiv m(f)$, i.e. the range of $f$ is naturally parametrized.
$\diamond$
\hfil\break

One can define the category of Einsteinian particles.\br

\noindent
{\tt 3.11. DEFINITION:}\hfil\break
For two points $a,b \in I$ we call $|a-b| \in |[s]|$  the proper time passed
between the two points $f(a),\  f(b)$ along the path of the particle $f$.
$\diamond$
\hfil\break

\noindent
{\tt 3.12. DEFINITION:}\hfil\break
Let $f$ be a particle. We call $p=f'$ the four-momentum of the particle,
\hfil\break
$v=(u\circ f)'= 1/m(f)\cdot p$  the four-velocity of the particle,\hfil\break
$F=f''$ the four-force acting on the particle, \hfil\break
$a= 1/m(f)\cdot f'' = (u\circ f)''$ the four-acceleration of the particle.
\hfil\break
Note, that $p = m(f)v$,\hfil
$$F= m(f)a,$$
and $||v|| = 1$, $<a, v> = 0$ holds. $||p|| = m(f)\cdot c$,
$||p||^2/m(f)=m(f)c^2$  is called the rest energy of $f$. $\diamond$
\hfil\break

\noindent
{\bf 5. Force field, equation of motion}\br

\noindent
{\tt 3.13. DEFINITION:}\hfil\break
We call a map $F : V\times M_0 \to M_0/([s]^2)$ which satisfies $<F(x,p), p>
= 0$ a force field.\hfil\break

A particle $f$ is said to be determined by the force field $F$ if \hfil
$$F\circ (f,f') = f''.$$
$\diamond$
\hfil\break

One can define the category of Einsteinian force fields.\br

{\bf
\leftline{6. Space of four-velocities}}
\vskip+0.1in 
The space of four-velocities is $V(1)=\{ $ the timelike lines containing 0 in
$M_0$ $\} \equiv \{ v\in M_0/[kgs]\ | \ ||v||=1\ \} $, which is (by
definition) the 3-dimensional hyperbolic space. \hfil\break

\noindent
{\tt 3.14. DEFINITION:}\hfil\break
Let us choose a spacelike 3-dimensional subspace in $M_0$, denoted by $E$.
This determines two orthogonal projections: $P_E$ onto $E$ and $P_E^{\perp}$
onto $E^{\perp}$. Using these projections we define the (bijective) map
(which we call Cayley map):\hfil
$$ \Gamma_E : M_0 \to E/|[kgs]|, \ \ v\mapsto P_E(v)/ |P_E^{\perp}(v)|, \ \ \
\Gamma_E(0)=0.$$

$\Gamma_E$ maps $V(1)$ into $E/|[kgs]|$, onto the open ball $\{ v \in
E/|[kgs]|\ |\ ||v|| < c\}$. The pair $(E/|[kgs]|, \Gamma_E)$ is known as the
Beltrami - Cayley - Klein model of the hyperbolic space. ($E/|[kgs]|$ is a
hyperbolic space with the Riemannian metric on it which is pushed forward
onto it from $V(1)$ by $\Gamma_E$). $\diamond$
\hfil\break

\noindent
{\tt NOTE: }
Clearly this model carries more structure than a hyperbolic space. $P_E$
defines another model of the hyperbolic space. The choice of $E$ also
determines certain boosts and the stabilizer subgroup of $E$ in the Lorentz
group.\br

{\bf
\leftline{7. Effects of fixing a reference frame}}
\vskip+0.1in 
The choice of an inertial reference frame also determines a Cayley map, and
when one speaks of three-velocities in special relativity then it is just the
image of a four-velocity by a Cayley map. On the other hand, three-momenta
are obtained from four-momenta by a $P_E$ (and the energy component is
obtained by the $P_E^{\perp}$).   \hfil\break
An inertial reference frame also determines an origin, units in $[kg]$ and
$[kgs]$, a parametrization of $\bar P$, a certain $SO(3)$ subgroup  and
certain boosts in $\bar P$ and a basis of the Lie-algebra of $P$. \br

If a mass $m$, and a position vector $x^{\mu}$ is given in some reference
frame, then the corresponding vector in $R^5\equiv R^4\times R$ is\hfil
$$(mx^{\mu},m).$$
The coordinate form of the points of $M$ is \hfil
$$(x^{\mu},1),$$
of the points of $V(1)$ is\hfil
$$(v^{\mu},0),$$
and the functions $m$, $mt$, $md$, $d$, $u$ have the following forms:\hfil
$$m : (mx^{\mu},m) \mapsto m,$$
$$mt : (mx^{\mu},0) \mapsto m\sqrt{(-1)x^{\mu}x_{\mu}},$$
$$md : (mx^{\mu},0) \mapsto m\sqrt{x^{\mu}x_{\mu}},$$
$$u : (mx^{\mu},m) \mapsto (x^{\mu},1).$$
If $f(t)=(mx^{\mu}(\tau ),m)$ is a particle, then\hfil
$$p(\tau )=(mx'^{\mu}(\tau ),0),$$
$$F(\tau )=(mx''^{\mu}(\tau ),0),$$
$$v(\tau )=(x'^{\mu}(\tau ),0),$$
$$a(\tau )=(x''^{\mu}(\tau ),0).$$
Furthermore\hfil 
$$P_E : (x,t) \mapsto (x,0),$$
$$\Gamma _E : (x,t) \mapsto x/t.$$

\vfill
\eject

{\bf \hfil IV. OLD RESULTS ABOUT THE ROLE OF MASS\hfil}\break

So far we defined the mass as something connected with the action of the
Galilei or Poincare group on a linear space. The mass as a quantity connected
to the Galilei or Poincare group can be introduced in other settings as well.
In this section we wish to display that our approach is in accordance with
those in the literature.\hfil\break

According to the result of Wigner and Bargmann the  free particles of quantum
physics can be brought into correspondence with the elements  of certain
irreducible ray representations of the Galilei and Poincare group. These
representations are naturally parametrized by two quantities: $m \in R_0^{+}$
and $s\in N/2$ (integers and half-integers) which are identified with the
mass (really rest energy in the relativistic case), and spin of the particle
(so mass and spin are parameters). (In the case of the Poincare group these
parameters are the eigenvalues of the two Casimir operators.)\br
The analogous result in classical Hamiltonian dynamics is that some of the
transitive symplectic representations of the Galilei/Poincare group can be
naturally parametrized by a positive real number and these representations
can be brought into correspondence with the free particles. The number is
again identified with the mass of the particle.\br
We recall now these theorems and then we formulate them within the framework
of our setting. This is done to show that if we formulate these theorems in
our setting, then the mass defined by us gets into the role of the quantity
which is usually called the mass in the context of these theorems.
\hfil\break

\noindent
{\tt 4.1. THEOREM: } (see [5]) Let $R^3\times R$ be equipped with the
Galilei space-time structure (in the standard way). The timelike lines (which
are the possible trajectories of a free particle ) in $R^3\times R$ form a
6-dimensional manifold which can be identified with $R^3\times R^3$. The
Galilei group acts on this manifold transitively. Introducing the  symplectic
form $[(v_1, q_1), (v_2, q_2)] \mapsto <mv_1, q_2>-<mv_2,q_1>,\ m\in R^{+}$
the representation of the Galilei group turns into a transitive symplectic
representation of the Galilei group. Two such representations with $m_1$ and
$m_2$ are equivalent if and only if $m_1=m_2$.\br
( By equivalence of two symplectic representations we mean the existence of
a $G^R$-equiva-riant diffeomorphism $\phi$ for which $\phi_{*}\omega_1 =
\omega_2$.) $\diamond$
\hfil\break

\noindent
{\tt 4.2. THEOREM:} (see [6]) Let $X_H$ be a Galilei invariant Hamiltonian
field on $T^{*}R^3$. Then there exists a unique constant $m_0>0$ so that
$X_H$ corresponds to a free particle of mass $m_0$. $\diamond$
\hfil\break

($X_H$ is called Galilei invariant, if there is an action of $G^R$ on
$T^{*}R^3$, realized by symplectic diffeomorphisms so that \hfil\break
the space translations are represented by $(p,x)\mapsto (p,x+a)$,\hfil\break
the rotations are represented by $(p,x)\mapsto (Rp,Rx)$,\hfil\break
time translations are generated by $X_H$.)\hfil\break

(A free particle of mass $m_0>0$ is defined by the Hamiltonian
$H=||x||^2/(2m_0)+const.$, which is Galilei invariant, if the velocity
transformations are represented by $(p,x)\mapsto (p-m_0v,x)$.\hfil\break

\noi
Relativistic case:\hfil\break

\noindent
{\tt 4.3. THEOREM:} (see [5]) Let us denote the space of timelike lines in
the Minkowski space $R^3_1 = R^3\times R$ by $M$. This is a 6-dimensional
manifold which can be identified with $V(1)\times R^3$. Using the fact that
this is a submanifold of $R^3_1\times R^3_1$ we can introduce the following
symplectic form: $[(v_1,q_1),(v_2,q_2)]\mapsto <v_1, mq_2>-<v_2, mq_1>$,
($<,>$ is the Lorentzian form) with which $M$ carries a transitive symplectic
representation of the Poincare group. Two such representations are equivalent
if and only if $m_1=m_2$. $\diamond$
\hfil\break

\noindent
{\tt 4.4. THEOREM: } (see [6]) Let $X_H$ be Poincare invariant. Then there
exists a unique constant $m$ so that $H(p,q) = \sqrt{m^2+||p||^2}+const.$
$\diamond$
\hfil\break

Poincare invariance means here that there is a symplectic representation of
the Poinca-re group on $T^{*}R^3$ so that the translations are represented by
$(x,p)\mapsto (x+a,p)$, the rotations are represented by $(x,p)\mapsto
(Rx,Rp)$ and the time translations are generated by $X_H$.\hfil\break

The following theorems are reformulated versions of the above ones:\hfil\break

Nonrelativistic case:\br

\noindent
{\tt 4.5. THEOREM:} Let us denote the set of the timelike lines in $M_m$,
$m\ne 0$ by ${\cal F}_m$. This is a 6-dimensional manifold on which $G$ acts
transitively. The map $\pi : {\cal F}_m \to V(1)$ which assigns to every
element of $\cal F$ its velocity makes ${\cal F}_m$ a bundle over $V(1)$.
Each fiber is an affine space over $E_0$. The tangent space $T_x{\cal F}_m$
is $(E_0/[kgs])\times E_0$  for all $x\in {\cal F}_m$. So ${\cal F}_m$ has
the canonical (generalized) symplectic form : $[(v_1, mq_1),(v_2, mq_2)]
\mapsto <v_1, mq_2> - <v_2, mq_1>$. With this symplectic structure   the
action of $G$ on ${\cal F}_m$ is symplectic.\hfil\break
(Generalized means not real but 1-dimensional vector space valued.)\hfil\break

Concerning the relationship between ${\cal F}$-s with various masses we can
say that the following diagram is commutative: see figure 5.\hfil\break
Here $g$ and $\tilde g$ are the action of an element of $\bar G$ on the
corresponding spaces. In particular, two symplectic representations with
$m_1$ and $m_2$ are equivalent iff $m_1=\pm m_2$. \hfil\break

Choose an inertial reference frame. This determines symplectic isomorphisms
between $(E_0/[kgs])\times E_0$ and  the ${\cal F}_m$-s.  (And pushes forward
the transitive symplectic representation of ${\cal F}_m$.) Let $H_m :
E_0/[kgs]\times E_0 \to [kg][m]^2/[s]^2,\ \ (v,mq)\mapsto 1/2\cdot m||v||^2.$
Then $X_{H_m}$ generates the time translations parallel to the fourth axis.
$\diamond$
\hfil\break

Relativistic case:\hfil\break

\noindent
{\tt 4.6. THEOREM: } Let us denote the set of timelike lines in $M_m$  by
${\cal F}_m$. Again there is a natural bundle structure $\pi : {\cal F}_m \to
V(1)$. The fiber of the bundle over each point $v$ of $V(1)$ is an affine
space over the orthogonal of $v$ in  $M_0$. $P$ acts on ${\cal F}_m$
transitively. The tangent space $T_x{\cal F}_m$ for all $x\in {\cal F}_m$ is
$(\pi (x)^{\perp})\times (\pi (x))^{\perp}\otimes [kgs])$ for all $x\in
{\cal F}_m$. So we can fix
the (generalized) symplectic structure $((v_1, mq_1),(v_2, mq_2)) \mapsto
<v_1, mq_2>-<v_2, mq_1>$ which makes the action of $P$ symplectic.\br
Concerning the relationship between ${\cal F}$-s with various $m$-s we can
say that the following diagram is commutative: see figure 5.\hfil\break
Here $g$ and $\tilde g$ are the action of en element of $\bar P$ on the
corresponding spaces. In particular, two symplectic representations with
$m_1$, $m_2$ are equivalent iff $m_1=\pm m_2$.\br

Now choose an inertial reference frame, let $x$ be the unit vector in $M_0$
parallel with the fourth (time) axis. This reference frame determines
symplectic(!) diffeomorphisms $\ell_m : {\cal F}_m \to (x^{\perp})\times
(x^{\perp}\otimes [kgs])$, where the symplectic structure on
$(x^{\perp})\times (x^{\perp}\otimes [kgs])$ is the canonical one and $x$ is
the direction of the fourth (time) axis. Thus $P$ acts on $(x^{\perp})\times
(x^{\perp}\otimes [kgs])$ transitively by symplectic transformations. Let
$X_{H_m} : (v,mq) \mapsto \sqrt{m^2c^4+m^2||v||^4}$. Then $X_{H_m}$ generates
the time translations parallel to the fourth (time) axis. $\diamond$
\hfil\break

(To prove that the $\ell_m$-s are symplectic take the coordinates determined
by the chosen inertial reference frame. Then it turns out, that ${\cal F}_m$
is isomorphic (as bundle and symplectic manifold) to $T^{*}V(1)$, and of
course $(x^{\perp})\times (x^{\perp}\otimes [kgs])$ is isomorphic to
$T^{*}(x^{\perp})$. Now, composed with these isomorphisms the $\ell_m$-s turn
into the cotangent map of the projection $P_E$, where $ E=(x)^{\perp}$).
$\diamond$ \hfil\break

{\vskip+0.3in

{\bf 
\leftline{Conclusion}}
\vskip+0.2in

Our discussion covers the content of Newton's first and second law. The
statement of the third and fourth law would be straightforward now.\br
\indent
In our setting the notion of mass, force and momentum, which belong to
dynamics conventionally, have become geometrical  ones. It is the force
field which can be regarded as a proper dynamical notion. We can say, that
Newton's first and second law is  a specification of certain geometrical
circumstances on the one hand, and (a part of) a definition of  dynamics on
the
other hand.\br
We can see, that in our formulation we didn't need orientations, which
displays the known fact that Newtonian and special relativistic
mechanics do not have much to do with the orientation of space, time and
mass (mathematically). It should be noted, however, that the presence of
particles or a force field can easily determine an orientation of the
mechanical spaces.\br
We can also  see that Newtonian mechanics does not contain any natural units
of length, time and mass. Units are brought into Newtonian mechanics by
force fields and particles. Other branches of physics do contain fundamental
dimensional constants, of course. Relativistic mechanics is an example, and
here the role of the velocity of light is well understood. \hfil\break
Finally, one should note, (and this is an important point) that while in the
old results the notion of mass is  associated with free particles, this is
not at all the case in our setting.\hfil

\vskip+0.3in 
{\bf  
\leftline{Acknowledgements}}  
\vskip+0.2in  

I thank K. T\'oth for many discussions, reading the manuscript    
carefully and proposing a number of modifications.  \hfil  
  
\vfill
\eject

{\bf
\centerline{APPENDIX}
\vskip+0.2in
\centerline{OPERATIONS WITH ONE-DIMENSIONAL LINEAR SPACES}}
\vskip+0.3in
\noindent
Every linear space which occurs in this section is meant to be real.
\hfil\break

\noindent
{\tt A.1. DEFINITION:} (Division)\hfil\break
Let W be an arbitrary linear space and D a
1-dimensional one.\hfil\break
We call $W/D :=Hom(D,W)\equiv W \otimes D^{*}$ the quotient of $W$ and $D$.
(Care must be taken with this notation, it can be mixed up with the notation
of quotient space.)
For $w \in W$, $d \in D$ we call the element $w/d$ of $Hom(D,V)$ the quotient
of $w$ and $d$, where $w/d$ is determined by the property $(w/d) (d) = w$.
$\diamond$
\hfil\break

By product of vector spaces we mean tensorial product. One easily verifies,
that the usual identities of multiplication and division hold for the
division and multiplication introduced now, i.e.:\hfil
$$ D\otimes (V/D) \equiv (V/D)\otimes D \equiv V,\ \ \
(V/A)/B\equiv V/(A\otimes B), \ \ \ etc.$$
Note, that the linear spaces obtained from $W$ by multiplication or division
by other 1-dimensional linear spaces are not canonically isomorphic, but the
isomorphisms between them are determined up to nonzero real numbers (up to
positive real numbers in the case when the 1-dimensional space is oriented).
Thus the notion parallelism is meaningful regarding the elements from these
linear spaces.\hfil\break

\noindent
{\tt A.2. DEFINITION:}\hfil\break
Let $W$ and $X$ be arbitrary linear spaces and $A$ an 1-dimensional one.
Given a linear map $L : W \to X$, this determines the linear maps $L/A\ :\
W/A \to X/A ,\ \ w/a \mapsto x/a$ and $L\otimes A\ :\ W\otimes A \to X\otimes
A,\ \ w\otimes a \mapsto x\otimes a$,  where $a$ is a nonzero element of $A$
and the defined maps are independent of it. $L/A$ and $L\otimes A$ will be
denoted by $L$, too, for the sake of brevity. $\diamond$
 \hfil\break

The above notion of quotient of linear spaces was first used in the monograph
[4] on the structure of space-time in which it is  stated that when we have
dimensional quantities rather than bare numbers in physics then we treat 
1-dimensional 
linear spaces in fact. Our treatment of the question of
dimensions differs from that in [4], in particular, regarding the question of
the role of dimensions in Newtonian mechanics and special relativity.
\hfil\break
\noindent
Given two 1-dimensional linear spaces $A$ and $B$, the spaces $A/B$ and
$A\otimes B$ are oriented if and only if $A$ and $B$ are both oriented.
(Similar statement is true  when the dimension of $A$ is odd, and $B$ may be
unoriented if the dimension of $A$ is even.) On the other hand, the even
powers of $A$ : $A^2, A^4, ...$ are oriented anyway, the odd powers are
oriented if and only if $A$ is oriented. \hfil\break

\noindent
{\tt A.3. DEFINITION:} \hfil\break
The factor space of a 1-dimensional linear space $D$ with respect to 
multiplication by $(-1)$ is the nonnegative part of a 1-dimensional oriented
linear space which we call the absolute value of $D$ and denote it by $|D|$.
We call the factor map $|\ | : D \to |D|,\ d \mapsto \{d,-d\}$  absolute
value function.  $\diamond$
\hfil\break

For $D$ being oriented and being canonically isomorphic to $|D|$ are the
same thing.\hfil\break
\noindent
We can now introduce arbitrary rational powers of a 1-dimensional real linear
space  and of the elements of it.\hfil\break

\noindent
{\tt A.4. DEFINITION:}\hfil\break
Let $A$ be an oriented 1-dimensional linear space. We call a pair $(B,i)$ of
an oriented 1-dimensional linear space $B$ and an orientation preserving
linear isomorphism $i$ between $B^n$ and $A$ an n-th root of $A$ (n is a
positive even number). As any two n-th root of $A$ are canonically
isomorphic, we speak of {\sl the} n-th root of $A$ and denote it by $\root
n\of A$ . We call  the map $p^{-1}\circ i^{-1}$ the  extraction of root,
where $i$ is the linear isomorphism between $({\root n\of A})^n$ and $A$, and
$p : a \mapsto a^n$.
For an unoriented 1-dimensional linear space $D$ we define the n-th root as
the n-th root of $|D|$. $\diamond$
 \hfil\break

(One can find explicit realization for the n-th roots of a one-dimensional
linear space.)\hfil\break

\noindent
The definition of a proper rational (i.e. non-integer) power of $D$ is now
clear.
\hfil\break

\vskip+0.4in
{\bf
\leftline{REFERENCES}}
\vskip+0.3in

\noi
[1] H, Weyl, Space-Time-Matter, Dover publ., 1922\br
[2] V. I. Arnold, Mathematical methods of classical mechanics, Springer,
1989\br
[3] A. Prastaro, Geometry of PDEs and Mechanics, World Scientific Publishing
Co. Pte. Ltd., 1996\br
[4] T. Matolcsi, A Concept of Mathematical Physics, Models for Spacetime,
Akad\'emiai Kiad\'o, Budapest, 1984 \br
[5] J. M. Souriau, Structure des systemes dynamique, Dunod, Paris, 1970\br
[6] R. Abraham, J. E. Marsden, Foundations of Mechanics, The
Benjamin/Cummings Publishing Co., Inc., 1978\hfil

\vfill
\eject

figure 1.\br

$$\diagram{ 
G & \lharr{}{} & {\cal BT}^4 & \lharr{}{} & {\cal T}^4 \cr 
\uvarr{}{} && \uvarr{}{} && \uvarr{}{} \cr 
SO{\cal BT}^3 & \lharr{}{} & {\cal BT}^3 & \lharr{}{} &{\cal T}^3 \cr 
&&&& \uvarr{}{} \cr 
&&&& e \cr 
}$$ 

figure 2.\br 

$$\diagram{ 
G&&&&\cr 
\dvarr{}{}&&&&\cr 
SO{\cal BT}^1&\rharr{}{}&SO{\cal T}^1&\rharr{}{}&{\cal T}^1\cr 
\dvarr{}{} && \dvarr{}{} && \dvarr{}{} \cr 
SO{\cal B} & \rharr{}{} & SO(g) & \rharr{}{} & e \cr 
}$$ 

figure 3.\br 

$$\diagram{ 
P & \lharr{}{} & {\cal T}^4 & \lharr{}{} & e \cr 
}$$ 

\vfill 
\eject 

figure 4.\br 

$$\diagram{ 
P & \rharr{}{} & {\cal L} & \lharr{}{} & e \cr 
}$$

figure 5.\br 

$$\diagram{ 
T^2{\cal F}_{m_1} & \rharr{Tg\ot Tg}{} & T^2{\cal F}_{m_2} \cr 
\dvarr{\omega_1}{} & & \dvarr{\omega_2}{} \cr 
[kgm][m]/[s] & \rharr{\tilde g}{} & [kgm][m]/[s] \cr 
}$$ 

\vfill
\bye
 \end